\documentclass{elsart}
\usepackage{graphicx}
\usepackage{graphics}
\usepackage{amsmath}

\newcommand{\bsigma}{\mbox{\boldmath $\sigma$}} 
\newcommand{\bsigmaa}{\mbox{\boldmath $\sigma'$}}
\newcommand{\btau}{\mbox{\boldmath $\tau$}} 

\newcommand{\bfeta}{\mbox{\boldmath $\eta$}}
\newcommand{\trace}{\mbox{Tr}}
\newcommand{\bold}[1]{\boldsymbol{#1}}
\newcommand{\trlog}{\ensuremath{\mbox{Tr\,log}}}
\newcommand{\bra}[1]{\ensuremath{\left( #1 \right)}}
\newcommand{\one}{{\boldsymbol{I}}}
\newcommand{\gauss}[1]{{\mathcal D} #1}

\newcommand{\bG}{\ensuremath{\mathbf{G}}}

\newcommand{\bR}{\ensuremath{\mathbf{R}}}
\newcommand{\bC}{\ensuremath{\mathbf{C}}}
\newcommand{\bD}{\ensuremath{\mathbf{D}}}

\newcommand{\nn}{\ensuremath{\eta}}

\newcommand{\av}[1]{\ensuremath{\left\langle #1 \right\rangle}}
\newcommand{\avs}[1]{\ensuremath{\left\langle \hspace*{-1.8mm} \left\langle 
                                         #1
				 \right\rangle \hspace*{-1.8mm} \right\rangle
				      }}
\newcommand{\bpsi}{\mbox{\boldmath ${\psi}$}}
\newcommand{\bbeta}{\mbox{\boldmath ${\eta}$}}

\begin{document}

\begin{frontmatter}

\title{Synchronous versus sequential updating in the three-state Ising neural network with variable dilution}
\author[leuven]{D. Boll\'e},
\ead{desire.bolle@fys.kuleuven.be}
\author[porto]{R. Erichsen Jr.}
\ead{rubem@if.ufrgs.br}
and
\author[leuven,brussel]{T.~Verbeiren},
\address[leuven]{Instituut voor Theoretische Fysica, Katholieke 
Universiteit Leuven, Celestijnenlaan 200 D, B-3001 Leuven, Belgium}
\address[porto]{Instituto de F{\'\i}sica, Universidade Federal do Rio
Grande do Sul, Caixa Postal 15051. 91501-970 Porto Alegre, RS, Brazil}
\address[brussel]{Guidance NV/SA, Buro \& Design Center,
Heysel Esplanade B62, B-1020 Brussels, Belgium}

\begin{abstract}
The three-state Ising neural network with synchronous updating and variable dilution is discussed starting from the appropriate Hamiltonians. The thermodynamic and retrieval properties are examined using replica mean-field theory. 
Capacity-temperature phase diagrams are derived for several values of the pattern activity and different gradations of dilution, and the information content is calculated. The results are compared with those for sequential updating. The effect of self-coupling is established. 
Also the dynamics is studied using the generating function technique for both synchronous and sequential updating. Typical flow diagrams for the overlap order parameter are presented. The differences with the signal-to-noise approach are outlined.

\vspace*{0.5cm} 
\vspace*{0.5cm} 
\end{abstract}
\end{frontmatter}

\newpage
\section{Introduction}
The dynamics and the storage and retrieval properties of multi-state attractor neural networks have been studied over some time now and numerous results are available (see, e.g., \cite{bollebook} and references therein). The majority of the results obtained on the storage and retrieval properties concern sequential updating of the neurons. 

Recently, it has been realized that synchronous updating of the spins in disordered systems can lead to different physics \cite{BB04}-\cite{Nreport}. For example, for binary spins, it is known that the phase diagram of the sequential and synchronous Little-Hopfield neural network \cite{L74,H82} in the replica-symmetric approximation are different \cite{FK88}, whereas the phase diagrams of the Sherrington-Kirkpatrick model \cite{SK72} are the same \cite{Nreport}. For the three-state Ising ferromagnet the same stationary solutions appear except for negative couplings, while for the Blume-Emery-Griffiths ferromagnet the phase diagram for synchronous updating is much richer \cite{BB04}.

For the three-state Ising neural network the possible different physics between sequential and synchronous updating has not yet been studied. Looking at the literature we see that using sequential updating the equilibrium properties of the $Q$-Ising model for the fully connected architecture have been studied in \cite{Rie90} and \cite{BRS94} making a replica symmetric ansatz.  The results for the extremely diluted architecture appeared  in
\cite{BCS00} and have been extended later to the whole dilution range in \cite{TE01}.
The layered architecture with variable dilution has been examined recently \cite{TE04}.
Nothing has been reported on the equilibrium properties for the recurrent architectures, however, when synchronous updating is used. On the other hand, concerning dynamics 
no calculations were done for the $Q$-Ising model with sequential updating. For synchronous updating the work of \cite{DGZ87} on the exactly solvable  extremely diluted asymmetric Little-Hopfield model has been extended to the $Q$-Ising model in \cite{Yed89}. 
Later on, the dynamics for the fully connected and the extremely diluted symmetric $Q$-Ising architecture have been solved in \cite{BJS98} and, respectively, \cite{BJS99}. The latter studies make use of the so-called signal-to-noise analysis (see, e.g., \cite{bollebook} for references on this method). An extension to the whole dilution range, in analogy with \cite{TE01} has not yet been given.

The aim of this work is precisely to fill the gaps mentioned above with a report on the study of the $Q=3$-Ising network with synchronous updating and variable dilution and a detailed comparison of the results obtained with the ones for sequential updating. First, the thermodynamic and retrieval properties are examined using replica symmetric mean-field theory. Capacity-temperature phase diagrams are derived for several values of the pattern activity and different gradations of dilution. Apart from the appearance of cycles the asymptotic behaviour is almost identical to the one for sequential updating. The spin-glass region is visibly enhanced, while the retrieval region, however, is only marginally enhanced. Only the addition of self-couplings can enlarge the retrieval region substantially, especially in the case of strong dilution. A calculaton of the information content shows that both for synchronous and sequential updating the three-state networks are robust against the interference of static noise coming from random dilution. Next, the dynamics of the model is studied using the generating function technique \cite{MSR73,Co1d}. As an illustration some typical flow diagrams for the overlap order parameter are presented. It is possible to extract the result for sequential updating from the one for synchronous updating. As in the Hopfield model \cite{BBV04} one can argue that the signal-to-noise analysis used before in the literature is a short memory approximation correct up to the third time step. And it can also be shown that the signal-to-noise analysis can be made exact.

The rest of the paper is organized as follows. In Section~2 the three-state Ising neural network with synchronous updating and variable dilution is introduced. Section~3 reports on the replica symmetric mean field theory calculation of the free energy and the fixed-point equations for the relevant order parameters. In Section~4 the phase diagrams and retrieval properties are discussed for arbitrary temperatures as a function of the gain parameter, the amount of dilution and the strength of the self-coupling. Section~5 discusses the dynamics for the model using the generating functional analysis and comments on the relation with the signal-to-noise analysis. In Section~6 some concluding remarks are given. Finally, the appendix presents the explicit saddle-point equations.

\section{The three-state Ising neural network}

Consider a network of N neurons, $\bsigma=\{\sigma_1,...,\sigma_N\}$,
which can take
values from the set $\mathcal{S}=\{-1, 0, 1\}$. In this network we want to store
$p=\alpha N$ patterns, $\{\xi_i^\mu\}$, $i = 1, \dots, N$ and $\mu = 1, \dots, p$.
They are supposed to be independent identically distributed random variables
(i.i.d.r.v.) with respect to $i$ and $\mu$, drawn from a probability distribution
given by
\begin{equation}
\textrm{P}(\xi_i^\mu) = a \, \delta(1 - (\xi_i^\mu)^2)
                     + (1-a) \, \delta(\xi^\mu_i)\, ,
\end{equation}
with $a$  the pattern activity defined by the expectation value 
\begin{equation}
\textrm{E}((\xi_i^\mu)^2) = a \ .
\end{equation}

The neurons are updated synchronously according to the transition probability 
\begin{eqnarray}
&&\mbox{W}\left[\bsigma(t+1)|\bsigmaa(t)\right]=
  \prod_{i=1}^N\mbox{Pr}\left(\sigma_i(t+1)=s\in \mathcal{S}|\bsigmaa(t) \right) \\
\label{three}
&&\mbox{Pr}\left(\sigma_i(t+1)=s\in\mathcal{S}|\bsigmaa(t) \right)=
     \frac{\exp[-\beta \epsilon_i(s|\bsigmaa(t))]}
          {\displaystyle{\sum_{s \in\mathcal{S}} 
                    \exp[-\beta \epsilon_i(s|\bsigmaa(t))] }}
\end{eqnarray}
with $\beta$ the inverse temperature and $\epsilon_i(s|\bsigma)$ an
effective single site energy function given by \cite{Rie90}
\begin{equation}
        \epsilon_i(s|{\bsigma})=
                -\left[\frac{1}{2}h_i({\bsigma})s-bs^2\right]
         \,, \label{eq:energy}
\end{equation}
where $b$ is the gain parameter of the system suppressing or enhancing the zero state of the neurons. The random local fields are defined by
\begin{equation}
h_{i}(\bsigma)=\sum_{j=1}^N J^c_{ij} \sigma_j \,  .
\end{equation}
The couplings $J^c_{ij}$ are taken to be of the form 
\begin{equation}
J_{ij}^c=\frac{c_{ij}}{c}J_{ij}\, ,
\label{coupdil}
\end{equation}
where the probability distribution of the $\{c_{ij}\}$ is given by
\begin{equation}
\mbox{P}(c_{ij})=c\delta(c_{ij}-1)+(1-c)\delta(c_{ij})\,.
\label{propdil}
\end{equation}
Hence, they allow for a diluted architecture. The  $J_{ij}$ are determined
via the Hebb rule
\begin{equation}
J_{ij} = \frac{1}{a N} \sum_{\mu=1}^p \xi^\mu_{i} \xi^\mu_{j} 
\label{couphebb}
\end{equation}

In order for the model to satisfy detailed balance (see,e.g., \cite{Co1d,P84}), the dilution has to be symmetric ($c_{ij}=c_{ji}$). In the case of extreme dilution, when  $c=0$, 
the average number of connections per neuron, $c N$, is still infinite
\begin{equation}
\lim_{N\rightarrow \infty} \frac{1}{c N}= 0\, .
\end{equation}

Finally, we recall that the detailed balance property for synchronous updating
is not destroyed by the presence of self-couplings, i.e., couplings of the form 
$J_{ii}^c$. Hence, we do allow for this type of couplings
and redefine $J_{ii}^c \rightarrow J_{0}J_{ii}^c= \alpha J_{0} $, with
$J_{0}$ a parameter and $\alpha=p/cN$ the capacity.

The long-time behaviour is governed by the Hamiltonian \cite{BB04}
\begin{equation}
 H(\bsigma)=-\frac{1}{\beta}\sum_{i=1}^N\ln{\left[\sum_{s\in\mathcal{S}}
                \exp{(\beta[h_i(\bsigma)s-bs^2])}\right]}
                 +b\sum_{i=1}^N\sigma_i^2 \, .
\label{parQham} 
\end{equation}
In addition, when evaluating traces over spins in the calculation of, e.g., the partition function, one realizes that the system is equivalent to one with a Hamiltonian involving a set of duplicate Ising spins (see, e.g., \cite{BB04,FK88,P84}), which can be written as 
\begin{equation}
\label{QIsyn}
H(\bsigma,\btau)=-\sum_{i,j}J_{ij}^c\sigma_i\tau_j 
      +b \sum_i[\sigma_i^2+\tau_i^2]
\end{equation}
such that $\underset{\bsigma}{\trace}\exp[ -\beta H(\bsigma)]= 
\underset{\bsigma}{\trace}\underset{\btau}{\trace}\exp[ -\beta H(\bsigma, \btau)] $. 

It is well-known that the equilibrium behaviour can be fixed-points and/or cycles of period 2, i.e., $\sigma_i(t)=\sigma_i(t+2), \forall i $. 

In the next Section we study the thermodynamic and retrieval properties of this model starting from the free energy.

\section{Replica mean-field theory}

In order to calculate the free energy we use the replica method as applied to dilute systems \cite{BCS00},\cite{TE01},\cite{WS91}-\cite{BCS76}. Since this method is really standard by now, at least for sequential updating, we refrain from giving any detailed calculations but concentrate on the main results and the differences between sequential and synchronous updating. Indeed, some complications arise due to the symmetry between the two types of spins in the Hamiltonian. 

Starting from the replicated partition function, performing the dilution average and the average over the condensed ($\mu=1$) and non-condensed ($\mu >1$) patterns we obtain for the replicated free energy density
\begin{eqnarray}
&&f_n  = 
         a \, \bold{m} \tilde{\bold{m}}
	  + \frac{\alpha c}{2\beta} 
	       \trlog \bra{\one - \beta \bold{A}}
\nonumber
\\          
&&	\hspace{1cm}  - (1-c) \frac{\alpha \beta}{2} 
	       \trace\bra{\bold{q} \bold{p}
	        + \bold{r}^2    }
          - \alpha (J_0 -c) \trace\bra{\bold{r}}
\nonumber
\\  
&&  \hspace{1cm}        + \frac{\alpha \beta }{2}
	      \trace \bra{
	            \bold{q} \hat{\bold{q}}
	            + \bold{p} \hat{\bold{p}}
	            + 2 \bold{r} \hat{\bold{r}^\dagger}
		    }
          - \frac{1}{\beta}\av{ 
	  \log \underset{\bsigma \btau}\trace \exp\bra{-\beta \tilde{H}(\bsigma,\btau)}}_\xi
	  \ ,
\label{eq:qisingp:fe_n}
\end{eqnarray}
with
\begin{equation}
\tilde H(\bsigma,\btau)
= 
	     - \xi (\bold{m}  \bsigma 
	             + \tilde{\bold{m}}  \btau) 
  - \frac{\alpha\beta}{2} \bra{
	             \bsigma \hat{\bold{q}} \bsigma^\dagger
	           +  \btau \hat{\bold{p}} \btau^\dagger
	           + 2 \bsigma \hat{\bold{r}} \btau^\dagger}
   - \beta b \bra{\bsigma^2 + \btau^2} \, .
\end{equation}
The symbol $\one$ denotes the unit matrix in replica space and 
\begin{equation}
\bold{A} = 
  \left(
  \begin{matrix}
     \bold{q}   &   i \bold{r}\\
     i \bold{s} &   - \bold{p}
  \end{matrix}
  \right)
\ .
\nonumber
\end{equation}
Hereby the usual order parameters are introduced as replica matrices ($n$ is the replica index) with elements
\begin{eqnarray}
&& m_{\alpha}^{\mu}   
   = \frac{1}{aN} \sum_{i} \xi_i^{\mu} \sigma^\alpha_i \,, \qquad\qquad\qquad
  \tilde{m}_{\alpha}^{\mu}
   = \frac{1}{aN} \sum_{i} \xi_i^{\mu} \tau^\alpha_i  \,, \\
&& q_{\alpha\beta}   
   = \frac{1}{N} \sum_{i} \sigma^\alpha_i \sigma^\beta_i \,,  \qquad\qquad\qquad
  p_{\alpha\beta}   
   = \frac{1}{N} \sum_{i} \tau^\alpha_i \tau^\beta_i \,, \\
&& r_{\alpha\beta}   
   = \frac{1}{N} \sum_{i} \sigma^\alpha_i \tau^\beta_i \,,  \qquad\qquad\qquad
  s_{\alpha\beta}   
   = \frac{1}{N} \sum_{i} \tau^\alpha_i \sigma^\beta_i \, .
   \label{eq:hopp:qrsp}
\end{eqnarray}
Their conjugate variables are denoted with a hat.
The free energy should be interpreted as being extremised with respect to $\bold{m}$,
$\bold{\hat{m}}$, $\bold{q}$, $\bold{p}$ and $\bold{r}$.  We remark that the
corresponding result for sequential dynamics (see, e.g., \cite{BRS94,BCS00,TE01}) is
recovered by assuming $\bsigma=\btau$ and rescaling the temperature with a factor 2.   
Furthermore, we notice that compared with sequential updating, the replicated free energy (\ref{eq:qisingp:fe_n}) is more involved. It is clear that a priori the
matrices appearing in this expression are not necessarily symmetric.

As mentioned in the introduction it is known that synchronous updating can lead to two-cycles as stationary solutions. Results for the Hopfield model \cite{SMK96,tonithesis} and the Blume-Emery-Griffiths model \cite{jordithesis,BB05} show that such cycles do not appear in the retrieval region of the corresponding phase diagrams. Therefore, cycles have been neglected in the replica calculation of the retrieval properties of these models. In analogy, for the three-state Ising model discussed here we make a similar ansatz. 
We neglect cycles in the sequel of the replica calculation and we assume that the two sets of spins behave in a completely symmetric way.    Consequently, $\bold{q}=\bold{p}$, $\bold{r}=\bold{r}^\dagger$, and similarly
for the conjugate variables.  The same symmetry applies to $\bold{m}$ and
$\tilde{\bold{m}}$.  The free energy then becomes
\begin{eqnarray}
&& f  =  
         a \bold{m}^2 
	  + \frac{\alpha c}{2\beta} 
	       \trlog\bra{(\one - \beta \bold{r})^2
	         - \beta^2 \bold{q}^2}
\nonumber
\\
 && \hspace{1cm}  
          - (1-c) \frac{\alpha \beta}{2} 
	    \trace\bra{ \bold{q}^2 + \bold{r}^2  }
          - \alpha (J_0 -c) \trace\bra{\bold{r}}
\nonumber
\\
 && \hspace{1cm}       
          + \alpha \beta
	       \trace\bra{
	               \bold{q} \hat{\bold{q}}
	               + \bold{r} \hat{\bold{r}}
		      }
	  -\frac{1}{\beta}
	     \av{\log \underset{\bsigma \btau}\trace \exp \bra{-\beta
	     \tilde{H}(\bsigma,\btau)}}_\xi \ , 
\end{eqnarray}
with
\begin{equation}
\tilde H(\bsigma,\tau)
= 
  -\xi \, \bold{m} (\bsigma + \btau) 
  - \frac{\alpha\beta}{2} \bra{
	             \bsigma \hat{\bold{q}} \bsigma^\dagger
	           +  \btau \hat{\bold{q}} \btau^\dagger
	           + 2 \bsigma \hat{\bold{r}} \btau^\dagger} 
                    - \beta b \bra{\bsigma^2 + \btau^2 } \,.
\end{equation}
 
Next, we take the replica symmetry (RS) ansatz
\begin{equation}
m_\alpha = m \ ,
\quad
q_{\alpha\beta} = q_1 \ ,
\quad
q_{\alpha\alpha} = q_0 \ ,
\quad
r_{\alpha \beta} = r_1 \ ,
\quad
r_{\alpha\alpha} = r \ , \quad \alpha\ne\beta
\end{equation}
and similarly for the conjugated variables. Due to the non-cycle ansatz, we have in addition 
\begin{equation}
r_1 = r_{\alpha\beta} = q_{\alpha\beta} = q_1    
\end{equation}
such that the RS free energy reads 
\begin{eqnarray}
&& f 
   =  a m^2
    + \frac{\alpha c}{2\beta} 
         \log\Big[ (1-\chi_r)^2 - \chi^2 \Big]
    - \alpha c \frac{q}{1-\chi_r - \chi}
  \nonumber \\
&& \hspace{1cm}
    - (1-c)\frac{\alpha \beta }{2} 
         \bra{ q_0^2 - 2 q^2 + r^2 }
    + {\alpha \beta} 
         \bra{ 
	       \hat q_0 q_0 - 2 \hat{q_1} q_1 + \hat{r} r 
         }
  \nonumber \\
 && \hspace{1cm} 
    + \alpha (c-J_0) r
    - \frac{1}{\beta} \av{ \int \gauss{z}
    \log \underset{{\sigma \tau}}\trace \exp\bra{-\beta \tilde H(\sigma,\tau|z)}  }_\xi \ ,
\label{eq:qising:fe-par}
\end{eqnarray}
with the effective Hamiltonian 
\begin{equation}
\tilde H(\sigma,\tau|z) 
     = -\bra{\xi m  + \sqrt{\alpha \hat{q}_1}z}
            (\sigma + \tau ) 
	   + \bra{b -  \frac{1}{2} \alpha \hat \chi} (\sigma^2 + \tau^2)
    	   - \alpha \hat \chi_r \sigma \tau
\label{eq:qising:ham-par}
\end{equation}
and $\gauss{z}$ the gaussian measure $ \gauss{z}= dz (2 \pi)^{-1/2} \exp(-z^2/2)$.
We have also defined the susceptibilities 
\begin{equation}
\chi = \beta(q_0-q_1) \ ,
\qquad
\chi_r = \beta(r-q_1) \ ,
\label{eq:qisingp:chichir}
\end{equation}
and their conjugate expressions
\begin{equation}
\hat \chi = \beta (\hat q_0 - \hat q_1) \ ,
\qquad
\hat \chi_r = \beta (\hat r - \hat q_1) \ .
\end{equation}

The phase structure of the network is then determined by the solution of the following set of saddle-point equations
\begin{eqnarray}
&& m  = \frac{1}{a}\av{\xi \int \gauss{z} \,  
           \av{\sigma}_z }_\xi \ ,
\qquad
      r  = \av{ \int \gauss{z} \,  
           \av{\sigma  \tau}_z }_\xi \ ,
\label{eq:pising:sp:rs:mr} 	   \\ 
&& q_0  = \av{ \int \gauss{z} \,  
           \av{\sigma^2}_z}_\xi \ ,
\qquad
       q_1  = \av{ \int \gauss{z} \,  
           \av{\sigma}_{z}^2}_\xi \ ,
\label{eq:pising:sp:rs:q01}
\end{eqnarray}
where the average $\av{\cdot}_z$ is defined with respect to the 
effective Hamiltonian (\ref{eq:qising:ham-par}) and
\begin{eqnarray}
&& \hat{q}_1 
     = q_1 \Big[
          (1-c) 
          + c \frac{1}{(1-\chi_r - \chi)^2} 
	\Big] \ ,
 \label{saddlecon0} \\
&& \hat \chi
    = (1-c) \chi  + c \frac{\chi}{(1-\chi_r)^2 - \chi^2} \ ,
    \label{saddlecon1}   \\
&& \hat \chi_r
   = (1-c) \chi_r  + c \frac{1-\chi_r}{(1-\chi_r)^2 - \chi^2}
      + (J-c)
\ .  \label{saddlecon}
\end{eqnarray}

Compared with sequential updating \cite{BRS94,BCS00,TE01} we notice the extra equations for $r$ and $\tilde{\chi}_r$
expressing that $\sigma$ and $\tau$ can be affected differently by thermal fluctuations and the fact that the equations for the conjugate parameters are different. In the appendix we present the explicit forms for (\ref{eq:pising:sp:rs:mr})-(\ref{eq:pising:sp:rs:q01}).

\section{Phase diagrams and retrieval properties}

First, we look at the special case of low loading ($\alpha=0$). In that case the effective
Hamiltonian  (\ref{eq:qising:ham-par}) simplifies to 
\begin{equation}
\tilde H(\sigma,\tau|z) 
     = -\xi m (\sigma + \tau ) 
	   + b (\sigma^2 + \tau^2)
\ ,
\end{equation}
which practically means that the two spin variables occurring in the
expression for the free energy (\ref{eq:qising:fe-par}) become independent yielding
$f_\text{par} = 2 f_\text{seq}$
as found before for other models \cite{BB04,AGS85,vH86}. Consequently, the saddle point equations are the same for synchronous and sequential updating and both yield the same stationary states.  An explicit calculation shows that the low loading results correspond to those of the $3$-Ising ferromagnet. For the relevant phase diagram we refer to \cite{BB04} (Fig.~2 for $J>0$).

Next, we turn to finite loading and solve the saddle point equations (\ref{eq:pising:sp:rs:mr})-(\ref{saddlecon}) numerically for arbitrary temperature $T$. 
We present the phase diagrams for some representative values of the
parameters $c$, $b$ and $J_0$. We focuss on uniformly distributed patterns
($a=2/3$). Results for sequential updating of the network are included for comparison.

\vspace*{1cm}
\begin{figure}[ht]
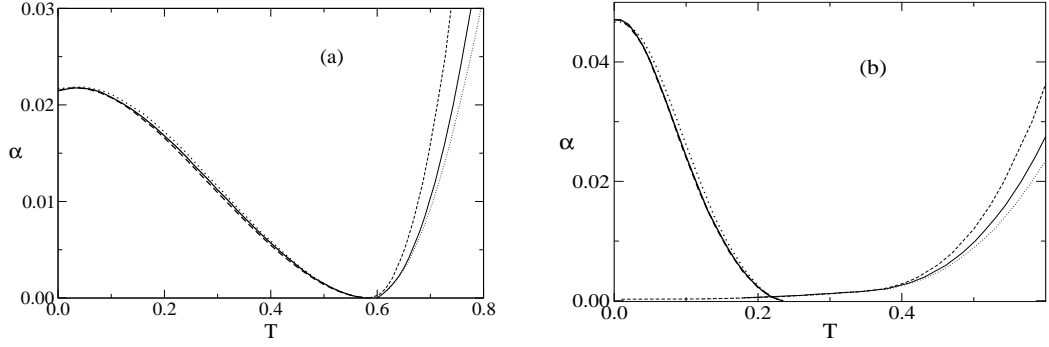

\begin{center}
\includegraphics[width=6.5cm,height=4.5cm]{bevsising1a.eps}\qquad 
\includegraphics[width=6.5cm,height=4.5cm]{bevsising1b.eps}
\caption{The RS $\alpha-T$ phase diagram for the 3-Ising network with $c=1$, $a=2/3$, $b=0.2$ (a) and $b=0.5$ (b). Solid (dotted) lines present synchronous updating for $J_0=0$ ($J_0=1.0$). Dashed lines are for sequential updating. The retrieval state is stable below the thick lines. The spin glass state is stable at the left of the thin lines.}
\label{fig1}
\end{center}
\end{figure}

In Figure 1 we present the RS $\alpha-T$ phase diagrams with ($J_0=1$) and without
($J_0=0$) self-coupling for the fully connectivity architecture ($c=1$) and gain parameter $b=0.2$ and $b=0.5$.
The retrieval transition is discontinuous in all cases. For a fixed $T$, the
retrieval capacity of the synchronously updated network with self-coupling is slightly 
larger than the one for the network without self-coupling, and both are slightly larger than the one for the sequentially updated network. However, the enhancement stays marginal, also for growing self-coupling. The spin glass transition is always continuous. For  $b=0.5$, contrary to $b=0.2$, the spin glass phase is not stable at small $\alpha$. But the retrieval phase is stable. This means that for $b=0.5$, small $\alpha$ and low $T$ the retrieval phase co-exists only with the paramagnetic phase, and not with the spin glass phase.

\vspace*{1cm}
\begin{figure}[ht]
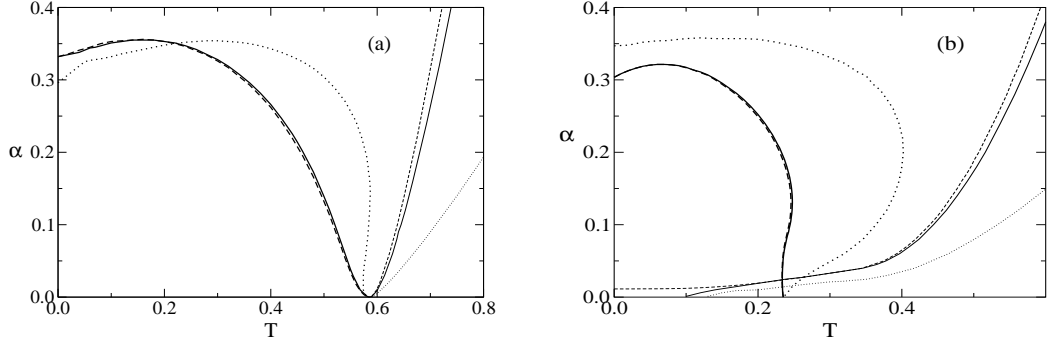

\begin{center}
\includegraphics[width=6.5cm,height=4.5cm]{bevsising2a.eps}\qquad 
\includegraphics[width=6.5cm,height=4.5cm]{bevsising2b.eps}
\caption{The same as Figure 1, but for the diluted network with $c=0.01$.}
\label{fig2}
\end{center}
\end{figure}

In Figure 2 we show the RS phase diagrams for the symmetrically diluted
networks with $c=0.01$. The other network parameters are kept identical to  those in
Figure 1. The retrieval (spin-glass) transitions remain discontinuous
(continuous). The enhancement of the retrieval capacity is again marginal for the network with synchronous updating and without self-coupling, compared to the one for sequential updating. However, increasing the self-coupling shows a substantial improvement, except for $\alpha=0$. Also the spin-glass region is visibly enlarged in that case. The remark concerning the stability of the spin-glass phase
for $b=0.5$ and low $T$ also applies to the diluted network, but with a small change: one
remarks (see the bottom left corner of Figure 2 (b)) that the spin-glass phase becomes 
stable at small $\alpha$ and low $T$, although there remains a region where only the retrieval and paramagnetic phases are stable. 
Finally, we notice a strong re-entrant behaviour in the retrieval region. This is related to the replica-symmetric approximation and is also seen in the Hopfield model \cite{WS91,CN92}. Consequently, we conjecture that the fact that for $b=0.5$, e.g., the maximal critical capacity for the network with self-coupling is larger compared to the one for the network without self-coupling, is an effect of this approximation. 

\vspace*{1cm}
\begin{figure}[ht]
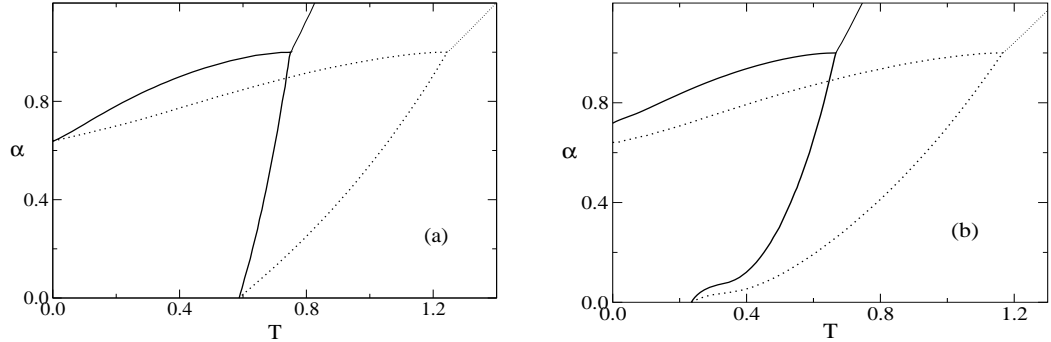

\begin{center}
\includegraphics[width=6.5cm,height=4.5cm]{bevsising3a.eps}\qquad 
\includegraphics[width=6.5cm,height=4.5cm]{bevsising3b.eps}
\caption{The same as Figure 1, but for the extremely diluted network with $c=0$. The curves for the sequentially and synchronously updated network without self coupling coincide.}
\label{fig3}
\end{center}
\end{figure}

The RS phase diagrams for the extremely diluted symmetric ($c=0$)
network are shown in Figure 3. The other parameters are kept identical
to those in Figures 1 and 2. The results for sequential and synchronous updating 
without self-coupling coincide. All transitions are continuous. Again, we notice a strong re-entrance behaviour in the retrieval region, in agreement with earlier results for this extremely diluted limit \cite{BCS00}. Increasing the  self-coupling makes the synchronously  updated network more robust against temperature, although, as expected, it 
has no effect on the critical temperature at zero $\alpha$. The re-entrance point is reached for the same value of $\alpha$.

\vspace*{1cm}
\begin{figure}[ht]
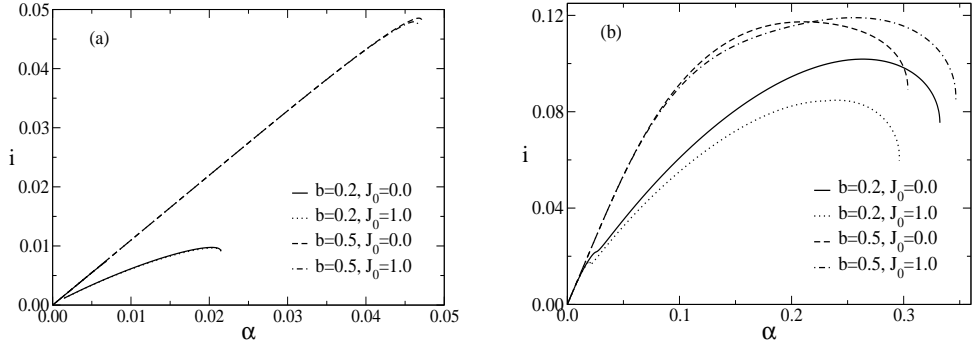

\begin{center}
\includegraphics[width=6cm,height=4.5cm]{bevsising4a.eps}\qquad
\includegraphics[width=6cm,height=4.5cm]{bevsising4b.eps}
\caption{ Information content $i$ as a function of the loading capacity
$\alpha$ for the $3$-Ising model with synchronous updating at $T=0$ with uniform
patterns ($a=2/3$), two values of $b$, with and without self-coupling and $c=1.0$ (a) and $0.01$ (b).}
\label{fig4}
\end{center}
\end{figure}

\vspace*{1cm}
\begin{figure}[ht]
\begin{center}
\includegraphics[angle=270,scale=0.30]{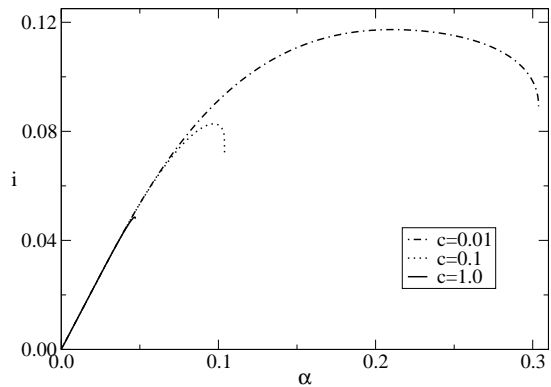}
\caption{ Information content $i$ as a function of the loading capacity
$\alpha$ for the $3$-Ising model with sequential updating, uniform
patterns, $b=0.5$, $T=0$, and $c=0.01$, $0.1$ and $1.0$.}
\label{fig5}
\end{center}
\end{figure}

Finally, we briefly study the robustness of these three-state networks
against the interference of static noise coming from random dilution. 
In Fig.~4 we show the information content of the model, being the product of the loading
capacity and the mutual information \cite{BDA00}, for synchronous updating and several amounts of dilution $c$ with and without self-coupling. 

We see that for the fully connected network ($c=1.0$, left figure), the results for $J_0=0$ and $J_0=1$ do practically coincide, in agreement with Fig.~1. 
We remark that for b=0.2, two solutions co-exist at small $\alpha$ ($0 \leq \alpha < 0.007$). One of them corresponds to perfect
retrieval, with $m=1.0$, $q_1=q_0=a, \chi=0$. The corresponding information content is  very close to that for $b=0.5$.
For the diluted network ($c=0.01$, right figure), in agreement with Fig.~2, self-coupling mostly leads to a higher information content for the optimal gain parameter $b=0.5$. (Again, for $b=0.2$ there exist two solutions for small $\alpha$ but the difference is very small and can hardly be seen on the scale of the figure.)

Next, by comparing with Fig.~5 for sequential updating, one notices that the results for sequential and synchronous updating (without self-coupling) are almost coincident. 
In all cases, we find that the  quality of the retrieval properties is affected very little, unless the amount of dilution is  high.

\section{Dynamics}

As mentioned in the introduction most studies in the literature on the dynamics of the $Q$-Ising model are based upon the signal-to-noise analysis. Except for the extremely diluted asymmetric and layered architectures, which can be solved exactly in closed form, the dynamics for the other architectures, only examined for synchronous updating, is obtained in the form of a recursive scheme. Recently, by a comparison with the generating functional analysis (GFA), it has been found \cite{BBV04} for the Hopfield model that the application of this method in the study of the dynamics involves a short-memory approximation implying that the results are only exact up to the third time step, although they stay very accurate in the retrieval region for further time steps, but not so in the spin-glass region. It has also been shown that the signal-to-noise analysis can be made exact, leading to the same results as the generating functional analysis.

Therefore, it is interesting to reconsider the dynamics for the $Q=3$-state Ising model using this generating functional technique, which was introduced in \cite{MSR73} to the field of statistical mechanics and, by now, is part of many textbooks.  We closely follow the derivation in \cite{BBV04}, extend it to variable dilution and indicate the differences for the model at hand. Both sequential and synchronous updating are discussed. In fact, the discussion is made for general $Q$. Since the method itself has become rather standard, we restrict ourselves to a presentation of the main arguments.

The idea of the GFA approach to study dynamics \cite{MSR73,Co1d}
is to look at the probability to find a certain microscopic path in time.
The basic tool to study the statistics of these paths is the generating
functional
\begin{equation}
Z[{\bpsi}] = \sum_{\bsigma(0), \dots, {\bsigma}(t)}
             P[{\bsigma}(0), \dots, {\bsigma}(t)]
    \, \prod_{i=1}^N \prod_{t'=0}^t e^{-i \, \psi_i(t') \, \sigma_i(t') }
    \label{functional}
\end{equation}
with $P[{\bsigma}(0), \dots, {\bsigma}(t)]$ the probability
to have a certain path in phase space
\begin{equation}
P[{\bsigma}(0), \dots, {\bsigma}(t)]
 = P[{\bsigma(0)}] \prod_{t'=0}^{t-1} W[{\bsigma}(t'+1)|{\bsigma}(t')]
\end{equation}
and $W[{\bsigma(t'+1)}|{\bsigma}(t')]$ the transition probabilities from
$\bsigma'$ to  $\bsigma$.  
Synchronous updating is expressed by
(recall eq.~(\ref{three})).
\begin{equation}
W[{\bsigma}(t'+1)|{\bsigma}(t')]= 
  \prod_{i=1}^N\frac{\exp{\left(\beta\sigma_i(t'+1)
        \epsilon_i(\sigma_i(t'+1)|\bsigma(t'))\right)}}
        {\underset{\sigma}{\textrm{Tr}}
     \exp{\left(\beta\sigma \epsilon_i(\sigma|\bsigma(t'))  \right)}} 
\end{equation}
where $\epsilon_i(\sigma_i(t'+1)|\bsigma(t'))$ now includes a time-dependent external field $\theta_i(s)$ in order to  define a response function  
\begin{equation}
\epsilon_i(\sigma_i(t'+1)|\bsigma(t'))= 
 -\left[\frac{1}{2}h_i({\bsigma(t')})\sigma(t'+1)-b(\sigma(t'+1))^2\right] +\theta_i(t')\,.
\end{equation}

One can find all the relevant order parameters, i.e., the overlap
$m(t)$, the correlation function $C(t,t')$ and the response function
$G(t,t')$, by calculating appropriate derivatives of the above functional (\ref{functional})
and letting ${\bpsi}=\{\psi_i\}$ tend to zero afterwards
\begin{eqnarray}
m(t)    & =& i \lim_{\bpsi\to0}
           \frac{1}{aN} \sum_{i}\xi_i
           \frac{\delta Z[\bpsi]}{\delta\psi_i(t)} \,,
	   \label{factora} \\
G(t,t') & =& i \lim_{\bpsi\to0}
           \frac{1}{N} \sum_{i}
           \frac{\delta^2 Z[\bpsi]}{\delta\psi_i(t) \delta\theta_i(t')} \,, \\
C(t,t') & =& - \lim_{\bpsi\to0}
           \frac{1}{N} \sum_{i}
           \frac{\delta^2 Z[\bpsi]}{\delta\psi_i(t) \delta\psi_i(t')}\,.
\end{eqnarray}

A difference with the Little-Hopfield model \cite{BBV04} is the presence of the factor $1/a$
in (\ref{factora}). First, due to the proper scaling of the couplings with this factor (recall eq.~(\ref{couphebb})) the average over the non-condensed patterns does not introduce an extra factor. Secondly, in the saddle point, terms containing this extra factor vanish. As a consequence, the factor $1/a$ does not appear explicitly in the expressions at all. The only point to keep in mind is the occurrence of the term proportional to the factor $b$ in the dynamics (recall eq.~(\ref{eq:energy})).

Another difference with respect to the treatment in \cite{BBV04} is that one needs to average over the dilution. This is done before averaging over the non-condensed patterns by using the fact that the diluted couplings $J_{ij}^c$ are of order
$\mathcal{O}((cN)^{-1/2})$ and that the 
$J_{ij}\sim \mathcal{O}(N^{-1/2})$. Noting that the
distribution for the  dilution eq.~(\ref{propdil}) is i.i.d.r.v. for $i<j$  and expanding the exponential containing  the $J_{ij}^c$ up to order $\mathcal{O}(N^{-3/2})$ makes this average then straightforward.  

In the thermodynamic limit one expects the physics of the problem to be
independent of the quenched disorder and, therefore, one is
interested in derivatives of $\overline{Z[{\bpsi}]}$, whereby the
overline denotes the average over this disorder, i.e., over all pattern
realizations.
This results in an effective single spin local field given by
\begin{equation}
   \label{eq:h}
    h(t) = \xi\,m(t) + \alpha \sum_{t'=0}^{t-1} R(t,t') \, \sigma(t')
         + \sqrt{\alpha} \nn(t)
\end{equation}
with $\eta(t)$ temporally correlated noise with zero mean and correlation
matrix 
\begin{equation}
  \bD = c(\one-\bG)^{-1} \bC (\one - \bG^\dagger)^{-1} + (1-c)\bC
\label{corrmatrix}
\end{equation}
and the retarded self-interactions
\begin{equation}
   \bR = c(\one-\bG)^{-1} + (1-c)\bG +(J_0-c)\one \, .
\label{selfmatrix}
\end{equation}
The order parameters defined above can be written as
\begin{eqnarray}
m(t)    & = & \left\langle\!\left\langle \av{ \xi \sigma(t) }_\star
                \right\rangle\!\right\rangle \,,
   \label{orderm} \\
C(t,t') & = & \left\langle\!\left\langle
                   \av{ \sigma(t) \, \sigma(t') }_\star
           \right\rangle\!\right\rangle \,,\\
G(t,t') & = & \avs{\av{ \frac{\partial \sigma(t)}{\partial \theta(t')}
                                     }_\star} \, .
 \label{eq:ordergfa}     
\end{eqnarray}
The average over the effective path measure $\av{\cdot}$ is given by
\begin{equation}
    \label{eq:noise:dist}
\av{f}_\star
=
  \mbox{Tr}_{\{\sigma(1),\ldots, \sigma(t) \}}
    \int d { \bfeta}
    \, P({\bfeta})
    \, P({\bsigma} | {\bfeta})
    \, f \ ,
  \end{equation}
where $d { \bfeta}= \prod_{t'} \eta(t')$ and with
\begin{eqnarray}
    \label{eq:s|eta}
&& P({\bsigma}|{\bfeta})
 = 
   \prod_{t'=0}^{t-1}
      \frac{\exp(\beta \sigma(t'+1) h(t')-\beta b \sigma^2(t'+1))}
            {\underset{\sigma}{\textrm{Tr}}\,
         \left(\exp(\beta \sigma(t'+1) h(t')-\beta b \sigma^2(t'+1))\right)}
   \\
&& P({\bfeta})
= 
     \frac{1}{\sqrt{\det(2\pi\bD)} }
 \exp
   \left (
     -\frac{1}{2} \sum_{t',t''=0}^{t-1} \eta(t') \, \bD^{-1}(t',t'') \, \eta(t'')
     \right ) \ .
     \label{eq:eta}
\end{eqnarray}
where $ h(t') = \xi\,m(t') + \alpha \sum_{p} R(t',p) \, \sigma(p) + \nn(t')$.
The average denoted by the double brackets in (\ref{eq:ordergfa}) is over the condensed pattern and initial conditions.
The set of eqs.~(\ref{eq:h})-(\ref{selfmatrix}) and (\ref{eq:noise:dist})-(\ref{eq:eta}) represent an exact dynamical scheme for the evolution of the network from which all relevant order parameters can be obtained at all time steps.

The order parameters are given by (\ref{orderm})-(\ref{eq:ordergfa}). To acquire a more intuitive expression for the response function we note that also
\begin{equation}
G(t,t')  =  \avs{\av{ \frac{\partial \sigma(t)}{\partial \eta(t')}
    }_\star} \,.
\end{equation}
We remark that $G(t',t'') = 0$ for $t'\le t''$ and $D(t',t'')=D(t'',t')$
and that for all $t'<t$
\begin{equation}
    \label{eq:G:T>0}
G(t,t') = \beta \left\langle\!\left\langle
     \av{\sigma(t)\left[\sigma(t'+1) - 
        \frac{ \underset{\sigma}{\textrm{Tr}} \,  \sigma    \exp{\left(\beta
        \epsilon_i(\sigma_i(t'+1)|\bsigma(t'))\right)}}
        {\underset{\sigma}{\textrm{Tr}}
     \exp{\left(\beta\sigma \epsilon_i(\sigma|\bsigma(t'))  \right)}}       
	      \right]}_\star
                 \right\rangle\!\right\rangle 
\end{equation}
where $h(t')$ appearing in $\epsilon_i$ is given by (\ref{eq:h}). 

For sequential updating  we have to start  from the stochastic process
\begin{equation}
\label{put}
p_{s+1}(\bsigma)=\sum_{\bsigma'}W_{s}[\bsigma;\bsigma']\,p_s(\bsigma')
\end{equation}
with $p_{s+1}(\bsigma)$ the probability to be in a state $\bsigma$ at
time $s+1$ and $W_{s}[\bsigma;\bsigma']$ given by
\begin{equation}
\label{put2}
W_{s}[\bsigma;\bsigma']=
  \frac{1}{N}\sum_i\left\{
  w_i(\bsigma)\delta_{\bsigma,\bsigma'}  
  +  w_i(F_i\bsigma)\delta_{\bsigma,G_i\bsigma'}+
  w_i(G_i\bsigma)\delta_{\bsigma,F_i\bsigma'}\right\} \,,
\end{equation}
with the shorthand 
$w_i(\bsigma)\equiv\mbox{Pr}\{\sigma_i(s+1)=\sigma_i|\bsigma(s)\}$ and where
$F_i$ and $G_i$ are cyclic spin-flip operators between the spin 
states. 
Each time step a randomly chosen spin is updated. In the thermodynamic limit the
dynamics becomes continuous because the characteristic time scale is $N^{-1}$.
The standard procedure is then to  update a random spin according to
(\ref{three}) and (\ref{eq:energy}) with time intervals $\Delta$ that are 
Poisson distributed with mean $N^{-1}$ . 
We can then write a continuous master equation in the thermodynamic limit
\begin{eqnarray}
\frac{d}{ds}p_s(\bsigma)&\equiv& \lim_{\Delta\rightarrow
  0}\frac{p_{s+\Delta}(\bsigma)-p_s(\bsigma)}{\Delta}  \\
&=&\sum_i\left\{
(w_i(\bsigma)-1)p_s(\bsigma)  
+   w_i(F_i\bsigma)p_s(F_i\bsigma)+
    w_i(G_i\bsigma)p_s(G_i\bsigma)\right\} \nonumber 
    \label{conp}
\end{eqnarray}

In that case the effective path average reads
\begin{equation}
    \label{eq:noise:distseq}
\av{f}_\star
=
    \int d { \bfeta}
    \, P({\bfeta})
    \, [f]_ {\bfeta}\ ,
  \end{equation}
where $ [f]_ {\bfeta}$ is an average over the (effective) stochastic process conditioned to the noise $\bfeta$ generated by the dynamics (\ref{conp}) and the distribution $P({\bfeta})$ is Gaussian with correlation matrix $\bD$
\begin{equation}
P({\bfeta})
= 
     \frac{1}{\sqrt{\det(2\pi\bD)} }
 \exp
   \left (
     -\frac{1}{2} \int \,dt' \,dt'' \eta(t') \, \bD^{-1}(t',t'') \, \eta(t'')
     \right ) \ .
     \label{eq:etaseq}
\end{equation}  
 
This result is very similar to the one for synchronous dynamics (recall eq. \ref{eq:eta}). The main reason is that one does not need the explicit form of the transition rates of the Markovian process in order to derive the effective path average. If only the initial conditions factorize over the site index $i$ then all sites become independent in the thermodynamic limit. In general, this is a characteristic feature of mean-field systems.  

Specialising this discussion to three Ising states $\{-1,0,+1\}$ we have that for synchronous updating the trace in eq. (\ref{eq:s|eta}) is given by
\begin{equation}
{\underset{\sigma}{\textrm{Tr}}\,
         \left(\exp(\beta \sigma(t'+1) h(t')-\beta b \sigma^2(t'+1)\right)}=
	 1+ 2 \cosh(\beta h(t'))\exp{(-\beta b)} \,.
\end{equation}
For sequential updating the  spin-flip operators in eq. (\ref{put2}) are defined by
\begin{eqnarray}
&&F_i\Phi(\bsigma)
    =\Phi(\sigma_1,...,\sigma_{i-1},\frac{-3\sigma_i^2-\sigma_i+2}{2},
       \sigma_{i+1},...,\sigma_N) \nonumber \\
&& G_i\Phi(\bsigma)= F_i(F_i\Phi(\bsigma)) \, .
\end{eqnarray}

We have solved the dynamics numerically using the Eissfeller-Opper method \cite{EO92}. The idea thereby is to replicate the system into $M$ copies of the effective spin and let each of them evolve independently in time according to their own stochastic path with its own noise variable. Averages over the effective path measure (recall eqs. (\ref{eq:noise:dist}) and (\ref{eq:noise:distseq})) are replaced by averages over the ensemble of copies for large $M$. We have taken $M=10^5$.  
\vspace*{1cm}
\begin{figure}[ht]
\begin{center}
\includegraphics[width=6cm,height=4.5cm]{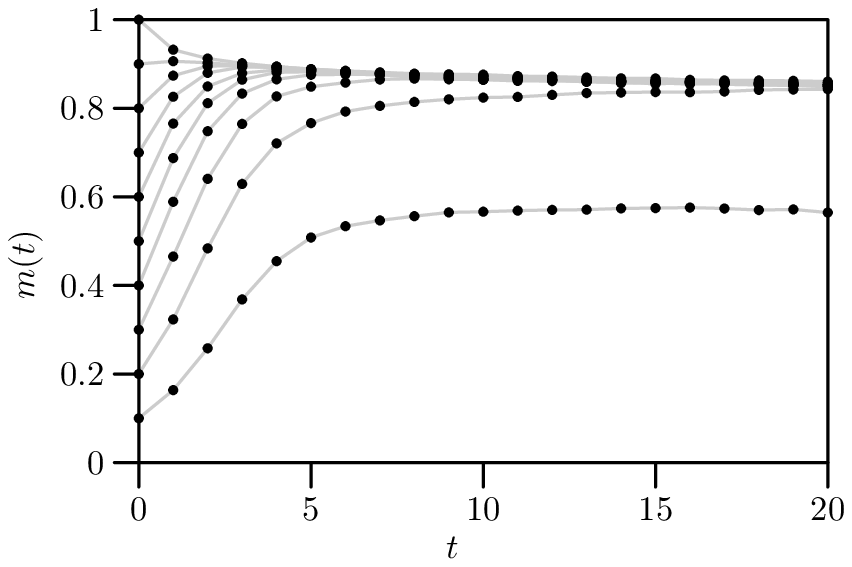}\qquad
\includegraphics[width=6cm,height=4.5cm]{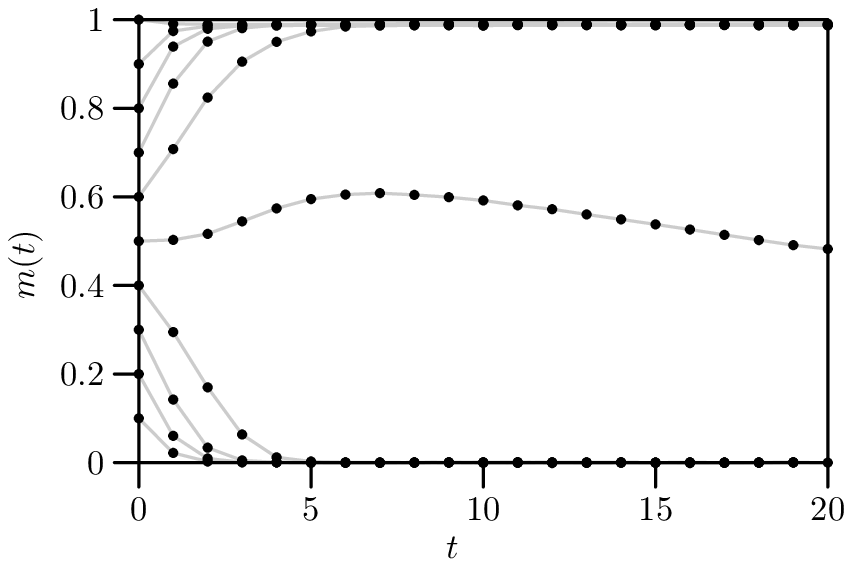}
\\
\includegraphics[width=6cm,height=4.5cm]{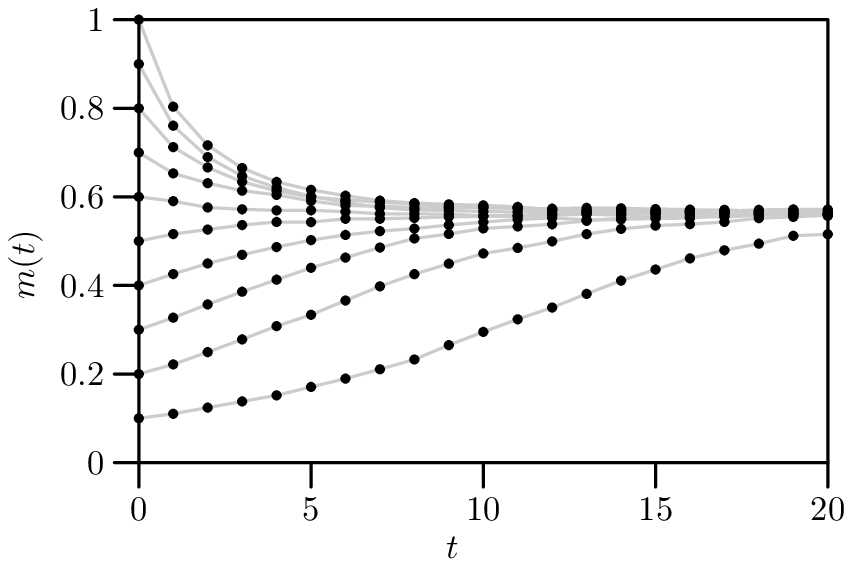}\qquad
\includegraphics[width=6cm,height=4.5cm]{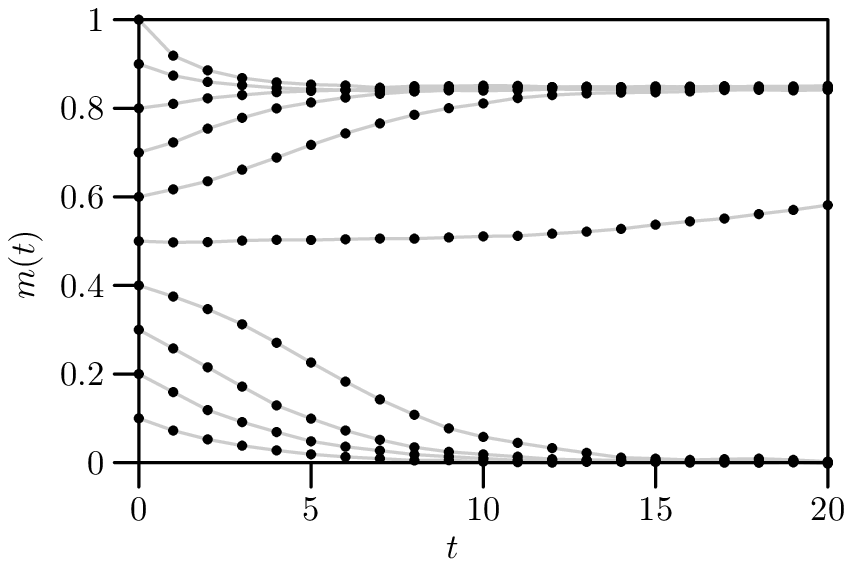}
\caption{ The overlap $m$ as a function of time using the Eissfeller-Opper algorithm with $10^5$ samples for the fully connected ($c=1$, top) and diluted ($c=0.01$, bottom) $3$-Ising model with synchronous updating and uniform patterns ($a=2/3$) for $\alpha=0.01$ and $J_0=0$. The other model parameters are   $b=0.2, T=0.3$ (top left), $b=0.5, T=0.1$ (top right), $b=0.2, T=0.5$ (bottom left) and $b=0.5, T=0.2$ (bottom right).  }
\label{figy}
\end{center}
\end{figure}

As an illustration we show some typical flow diagrams for the overlap order parameter in Figure 6 for some fully connected ($c=1$, top figures) and diluted ($c=0.01$, bottom figures) models. The model parameters are  $\alpha=0.01$ and $b=0.2, T=0.3$ (top left), $b=0.5, T=0.1$ (top right), $b=0.2, T=0.5$ (bottom left) and $b=0.5, T=0.2$ (bottom right). Self-couplings are set to zero.
 
Several remarks are in order. All situations shown are in the retrieval phase of the corresponding phase diagrams Figures 1 and 2. They all converge to the corresponding RS equilibrium results provided the initial overlap order parameter is large enough. This gives us also an idea of the basins of attraction. These basins are bigger for $b=0.2$ since the relevant model parameters are chosen deeper in the retrieval region, i.e.,  closer to the line of thermodynamic stability.

For completeness we end this Section on the discussion of the dynamics employing the GFA approach with a couple of remarks. 
First, this technique can also be applied to asymmetric dilution ($c_{ij}\neq c_{ji}$) and we then get for the correlation matrix and retarded self-interactions  
\begin{eqnarray}
\bR & =& c \, (\one - \bG)^{-1} + \Gamma (1-c) \, \bG
           + (J_0 - c) \one
\ ,
\label{eq:qising:RD:dil:R}
\\
\bD & =& c \, (\one - \bG)^{-1} \bC (\one - \bG^\dagger)^{-1} 
           + (1-c) \, \bC
\ .
\label{eq:qising:RD:dil:D}
\end{eqnarray}
with the asymmetry factor $\Gamma$ given by
\begin{equation}
 \Gamma=\frac{\left\langle c_{ij}c_{ji}\right\rangle -c^2}{c(1-c)}\, .
\end{equation}

Secondly, one can compare the GFA technique with the signal-to-noise analysis for the three-state Ising network \cite{BRS94,BCS00}. This comparison is formally completely analogous to the one for the Hopfield model discussed in detail in \cite{BBV04}. The only point at which one has to be careful is in deriving the recursion relations for the coloured noise $\bbeta$.  The diagonal of $\bC$, namely, is no longer equal to 1
but depends on time. Taking this into account one can take the results for the Hopfield model and extend it to $Q>2$-Ising models. The outcome is that, even for general $Q$, the signal-to-noise analysis is a short memory approximation to the real effective dynamics giving  results correct up to the third time step. Like for the other models, this approximation is extremely good in the retrieval region but fails in the spin glass region.
It is then a rather straightforward exercise to extend the revisited signal-to-noise analysis, proposed in
\cite{BBV04} to the $Q$-Ising case. Since the final results are completely equivalent to those of the GFA eqs.~(\ref{eq:h})-(\ref{selfmatrix}), we refrain from repeating further explicit details and refer to the work mentioned above (\cite{BBV04,tonithesis}).

\section{Concluding remarks}

We have studied the statics and dynamics of the three-state Ising neural network model with synchronous updating of the neurons and variable dilution.  We have followed as closely as
possible the discussion for the Hopfield model adapting the methods to the multi-state nature of the neurons.  Although the equations that describe both statics and dynamics are
formally rather similar to those of the Hopfield model, we have seen that the
physics behind it is quite different.

We have examined the thermodynamic and retrieval properties for this model using replica symmetric mean-field theory and have made a detailed comparison of the results with those for sequential updating.  
Capacity-temperature phase diagrams are derived for several values of the pattern activity and different gradations of dilution.
Apart from the appearance of cycles the asymptotic behaviour is almost identical to the one for sequential updating. The spin-glass region is visibly enhanced, while the retrieval region, however, is only marginally enhanced. Only the addition of self-couplings can enlarge the retrieval region substantially, especially in the case of strong dilution.
A calculation of the information content shows that both for synchronous and sequential updating the three-state networks are robust against the interference of static noise coming from random dilution.

Concerning the dynamics, we have extended the generating functional analysis to the study of $Q$-state spins.  As an illustration some typical flow diagrams for the overlap order parameter are presented in the case $Q=3$. It is possible to extract the result for sequential updating from the one for synchronous updating. A comparison with the signal-to-noise approach is made.
As for the Hopfield model one can argue that the signal-to-noise analysis used before in the literature is a short memory approximation correct up to the third time step. It can also be shown that this signal-to-noise analysis can be made exact.

\section*{Acknowledgments}

The authors would like to thank Jordi Busquets Blanco, Gyoung Moo Shim and Walter K. Theumann for useful discussions. R. E. thanks the kind hospitality and the support of the Instituut voor Theoretische Fysica, KULeuven. This work has been supported in part by the Fund of Scientific Research, Flanders-Belgium.

\section*{Appendix}

We write down explicitly the saddle-point equations discussed in Section 3. Therefore 
we need the  averages  $\left\langle\sigma\right\rangle_z$, $\left\langle\sigma^2\right\rangle_z$ and $\left\langle\sigma\tau\right\rangle_z$ with respect to the effective Hamiltonian (\ref{eq:qising:ham-par})
\begin{equation}
  \left\langle\sigma\right\rangle_z
  =\frac{2}{\mathcal{Z}}\left\{\sinh(2\beta M)\,
    {\rm e}^{\beta\left[2\left(-b+\frac{\alpha}{2}\hat{\chi}\right)
        +\alpha\hat{\chi}_r\right]}
    + \sinh(\beta M)\,{\rm
  e}^{\beta\left[-b+\frac{\alpha}{2}\hat{\chi}\right]}
\right\}\,,
  \label{eq19}
\end{equation}
\begin{equation}
  \left\langle\sigma\tau\right\rangle_z
  =\frac{2}{\mathcal{Z}}\left\{\cosh(2\beta M)\,
    {\rm e}^{\beta\left[2\left(-b+\frac{\alpha}{2}\hat{\chi}\right)
        +\alpha\hat{\chi}_r\right]}
    - {\rm e}^{\beta\left[2\left(-b+\frac{\alpha}{2}\hat{\chi}\right)
        -\alpha\hat{\chi}_r\right]}
  \right\}\,,
  \label{eq20}
\end{equation}
\begin{eqnarray}
  \left\langle\sigma^2\right\rangle_z
  &=&\frac{2}{\mathcal{Z}}\left\{\cosh(2\beta M)\,
    {\rm e}^{\beta\left[2\left(-b+\frac{\alpha}{2}\hat{\chi}\right)
        +\alpha\hat{\chi}_r\right]}
	\right.
  \nonumber \\
  &+&\left.
    \cosh(\beta M)\,{\rm e}^{\beta\left[-b+\frac{\alpha}{2}\hat{\chi}\right]}
    +{\rm e}^{\beta\left[2\left(-b+\frac{\alpha}{2}\hat{\chi}\right)
        -\alpha\hat{\chi}_r\right]}  
  \right\}\,,
  \label{eq21}
\end{eqnarray}
with
\begin{equation}
  M=\xi m + \sqrt{\alpha q_1} z\,.
  \label{eq17}
\end{equation}
Using the symmetry properties of these functions, we 
average over $\xi$ to obtain the saddle-point equations for the
retrieval state of one pattern at finite temperature. We get
\begin{equation}
  m=\int\mathcal{D}z\left\langle\sigma\right\rangle_z(m+\sqrt{\alpha\hat{q}_1}z)\,,
  \label{eq28}
\end{equation}
\begin{equation}
  r=a\int\mathcal{D}z\left\langle\sigma\tau\right\rangle_z (m+\sqrt{\alpha\hat{q}_1}z)
  +(1-a)\int\mathcal{D}z\left\langle\sigma\tau\right\rangle_z(\sqrt{\alpha\hat{q}_1}z)\,,
  \label{eq29}
\end{equation}
\begin{equation}
  q_0=a\int\mathcal{D}z\left\langle\sigma^2\right\rangle_z(m+\sqrt{\alpha\hat{q}_1}z)
  +(1-a)\int\mathcal{D}z\left\langle\sigma^2\right\rangle_z(\sqrt{\alpha\hat{q}_1}z)\,,
  \label{eq30}
\end{equation}
\begin{equation}
 q_1=a\int\mathcal{D}z\left[\left\langle\sigma\right\rangle_z\right]^2
  (m+\sqrt{\alpha\hat{q}_1}z)
  +(1-a)\int\mathcal{D}z\left[\left\langle\sigma\right\rangle_z\right]^2
  (\sqrt{\alpha\hat{q}_1}z)\,,
  \label{eq31}
\end{equation}
that need to be solved together with the algebraic equations (\ref{saddlecon0})-(\ref{saddlecon}).

\end{document}